\documentclass[twocolumn,floatfix,aps,superscriptaddress,eqsecnum]{revtex4}
\usepackage{graphicx}
\usepackage{amsmath}
\usepackage{amssymb}
\usepackage{bm}

\usepackage{color}

\begin{document}

\title{Universal chiral magnetic effect in the vortex lattice of a Weyl superconductor}
\author{M. J. Pacholski}
\affiliation{Instituut-Lorentz, Universiteit Leiden, P.O. Box 9506, 2300 RA Leiden, The Netherlands}
\author{C. W. J. Beenakker}
\affiliation{Instituut-Lorentz, Universiteit Leiden, P.O. Box 9506, 2300 RA Leiden, The Netherlands}
\author{\.{I}. Adagideli}
\affiliation{Faculty of Engineering and Natural Sciences, Sabanc\i\ University, 34956 Orhanl\i-Tuzla, Turkey}
\date{November 2019}
\begin{abstract}
It was shown recently that Weyl fermions in a superconducting vortex lattice can condense into Landau levels. Here we study the chiral magnetic effect in the lowest Landau level: The appearance of an equilibrium current $I$ along the lines of magnetic flux $\Phi$, due to an imbalance between Weyl fermions of opposite chirality. A universal contribution $dI/d\Phi=(e/h)^2\mu$ (at equilibrium chemical potential $\mu$ relative to the Weyl point) appears when quasiparticles of one of the two chiralities are confined in vortex cores. The confined states are charge-neutral Majorana fermions.
\end{abstract}
\maketitle

\section{Introduction}
\label{intro}

This paper combines two topics of recent research on Weyl fermions in condensed matter. The first topic is the search for the chiral magnetic effect in equilibrium \cite{Vaz13,Che13,Zho13,Ma15,Ala16, Zho16a,Bai16a,Zub16,Vol17,OBr17,Men19}. The second topic is the search for Landau levels in a superconducting vortex lattice \cite{Mas17,Liu17,Nic17,Pac18}. What we will show is that the lowest Landau level in the Abrikosov vortex lattice of a Weyl superconductor supports the equilibrium chiral magnetic effect at the universal limit of $(e/h)^2$, unaffected by any renormalization of the quasiparticle charge by the superconducting order parameter. Let us introduce these two topics separately and show how they come together.

The first topic, the chiral magnetic effect (CME) in a Weyl semimetal, is the appearance of an electrical current $I$ along lines of magnetic flux $\Phi$, in response to a chemical potential difference $\mu_{+}-\mu_{-}$ between Weyl fermions of opposite chirality. The universal value \cite{Nie83,Kha14,Bur15}
\begin{equation}
\frac{dI}{d\Phi}=\frac{e^2}{h^2}(\mu_{+}-\mu_{-})\label{ImuLmuR}
\end{equation}
follows directly from the product of the degeneracy $(e/h)\Phi$ of the lowest Landau level and the current per mode of $(e/h)(\mu_{+}-\mu_{-})$. A Weyl semimetal in equilibrium must have $\mu_{+}=\mu_{-}$, hence a vanishing chiral magnetic effect --- in accord with a classic result of Levitov, Nazarov, and Eliashberg \cite{Lev85,Naz86} that the combination of Onsager symmetry and gauge invariance forbids a linear relation between electrical current and magnetic field in equilibrium.

Because superconductivity breaks gauge invariance, a Weyl superconductor is not so constrained: As demonstrated in Ref.\ \onlinecite{OBr17}, one of the two chiralities can be gapped out by the superconducting order parameter. When a magnetic flux $\Phi$ penetrates uniformly through a thin film (no vortices), an equilibrium current
\begin{equation}
\frac{dI}{d\Phi}=\pm\frac{ee^\ast}{h^2}\mu_\pm\label{Imueq}
\end{equation}
appears along the flux lines, of a magnitude set by the equilibrium chemical potential $\mu_\pm$ of the ungapped chirality. The renormalized charge $e^\ast<e$ determines the degeneracy $(e^\ast/h)\Phi$ of the lowest Landau level in the superconducting thin film.

The second topic, the search for Landau levels in an Abrikosov vortex lattice, goes back to the discovery of massless Dirac fermions in \textit{d}-wave superconductors \cite{Gor98,And98}. In that context scattering by the vortex lattice obscures the Landau level quantization \cite{Mel99,Fra00,Mar00}, however, as discovered recently \cite{Pac18}, the chirality of Weyl fermions protects the zeroth Landau level by means of a topological index theorem. The same index theorem enforces the $(e/h)\Phi$ degeneracy of the Landau level, even though the charge of the quasiparticles is renormalized to $e^\ast<e$. Does this topological protection extend to the equilibrium chiral magnetic effect, so that we can realize Eq.\ \eqref{Imueq} with $e^\ast$ replaced by $e$? That is the question we set out to answer in this work.

The outline of the paper is as follows. In the next section we formulate the problem of a Weyl superconductor in a vortex lattice. We then show in Sec.\ \ref{sec_chiralitylocalization} that a flux bias of the superconductor can drive the quasiparticles into a topologically distinct phase where one chirality is exponentially confined to the vortex cores. The unconfined Landau bands contain electron-like or hole-like Weyl fermions, while the vortex-core bands are charge-neutral Majorana fermions. The consequences of this topological phase transition for the chiral magnetic effect are presented in Sec.\ \ref{sec_CME}. We support our analytical calculations with numerical simulations and conclude in Sec.\ \ref{sec:conclude}.

\section{Formulation of the problem}
\label{sec_formulation}

\begin{figure}[tb]
\centerline{\includegraphics[width=0.6\linewidth]{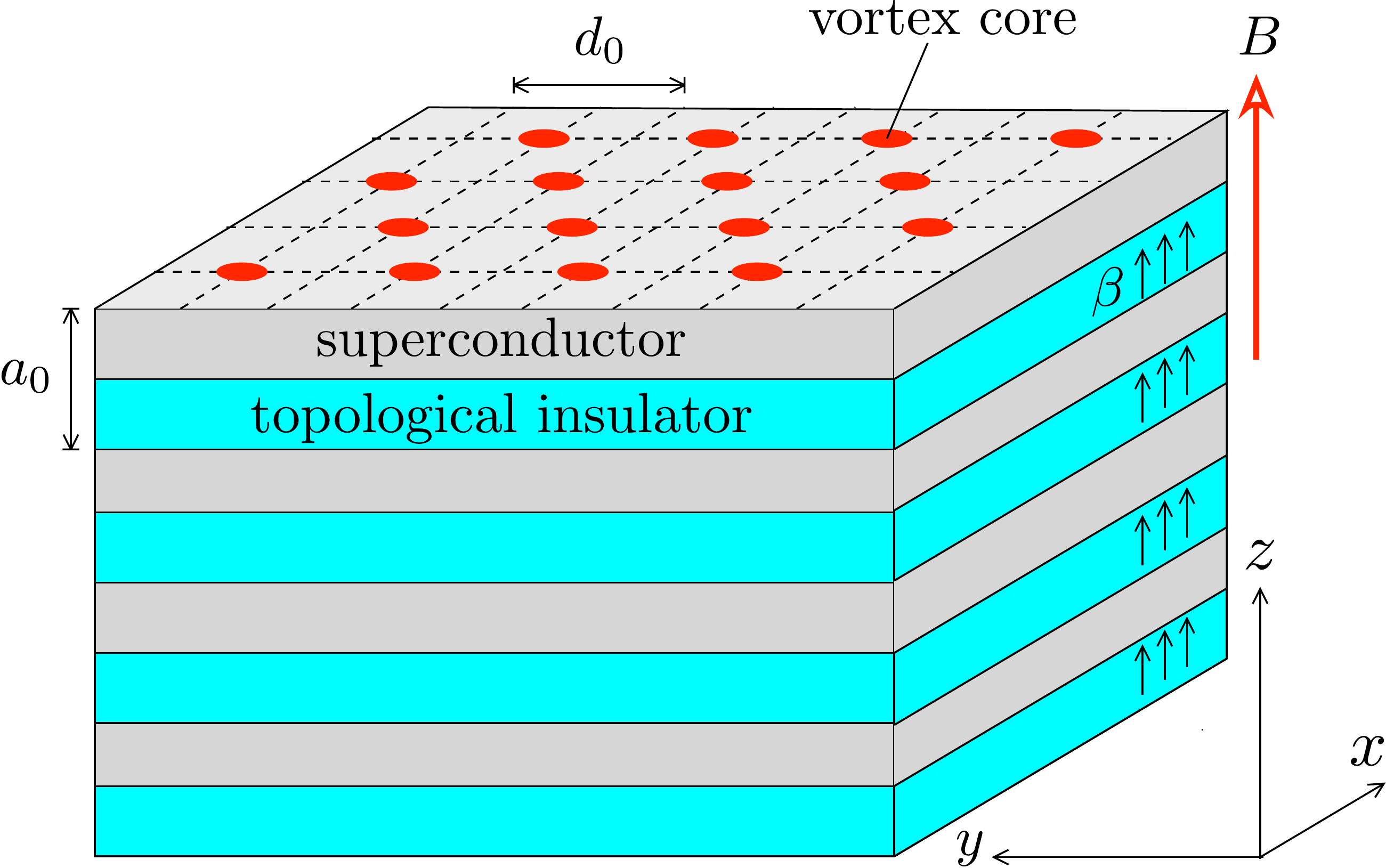}}
\caption{Cross-section through a heterostructure of alternating topological insulator layers and superconducting spacer layers. A perpendicular magnetization $\beta$ separates a pair of Weyl cones of opposite chirality along $k_z$. Each Weyl cone is twofold degenerate in the electron-hole degree of freedom, mixed by the superconducting pair potential $\Delta_0$. The mixing leaves the Weyl cones gapless, as long as the pair potential $\Delta_0$ remains smaller than $\beta$.
}
\label{fig_hetero}
\end{figure}

We consider a multilayer heterostructure, see Fig.\ \ref{fig_hetero}, composed of layers in the $x$--$y$ plane of a magnetically doped topological insulator (such as $\text{Bi}_2\text{Se}_3$), separated in the $z$-direction by a normal-insulator spacer layer. The tight-binding Hamiltonian is \cite{Bur11}
\begin{align}
H_0(\bm k) ={}& \sum_{i=x,y,z} \tau_z\sigma_i\,t_i\sin k_i a_i+ \beta\tau_0\sigma_z\nonumber\\
& + \tau_x\sigma_0\sum_{i=x,y,z}t'_i(1-\cos k_i a_i )  - \mu\tau_0\sigma_0,\label{eq:tb_normal}
\end{align}
where $t_i$, $t'_i$ are nearest-neighbor hopping energies, $a_i$ are lattice constants, and $\mu$ is the chemical potential. For simplicity we will equate $a_i=a_0$ and $t_i=t'_i=t_0$ for $i=x,y,z$.

The Pauli matrices $\sigma_i$ ($i=x,y,z$, with $i=0$ for the unit matrix) act on the spin degree of freedom of the surface electrons in the topological insulator layers. The $\tau_z=\pm 1$ index distinguishes the orbitals on the top and bottom surfaces. Magnetic impurities in the topological insulator layers produce a perpendicular magnetization, leading to an exchange splitting $\beta$. A Weyl point with a linear dispersion appears at $\bm{k}=(0,0,\pm\beta/a_0 t_0)$. For ease of notation we will set $a_0$, $t_0$, and $\hbar$ to unity.

Following Meng and Balents \cite{Men12}, the spacer layer may have a spin-singlet \textit{s}-wave pair potential $\Delta=\Delta_0 e^{i\phi}$. The pair potential induces superconductivity in the top and bottom surfaces of the topological insulator layers, as described by the Bogoliubov-De Gennes Hamiltonian
\begin{subequations}
\label{Hamiltonian_def}
\begin{align}
&{\mathcal{H}}(\bm k) = \begin{pmatrix}
H_0(\bm k - e\bm A) & \Delta_0 e^{i\phi} \\
\Delta_0 e^{-i\phi} & -\sigma_y H_0^\ast(-\bm k- e\bm A) \sigma_y
\end{pmatrix}.\label{eq:H_BdG}
\end{align}
\end{subequations}
We have introduced a vector potential $\bm{A}$ and take the electron charge $e>0$. For definiteness we also fix the sign $\beta>0$. The Fermi velocity $v_{\rm F}=a_0t_0/\hbar$ is unity for our chosen units.

\begin{figure}[tb]
\centerline{\includegraphics[width=0.9\linewidth]{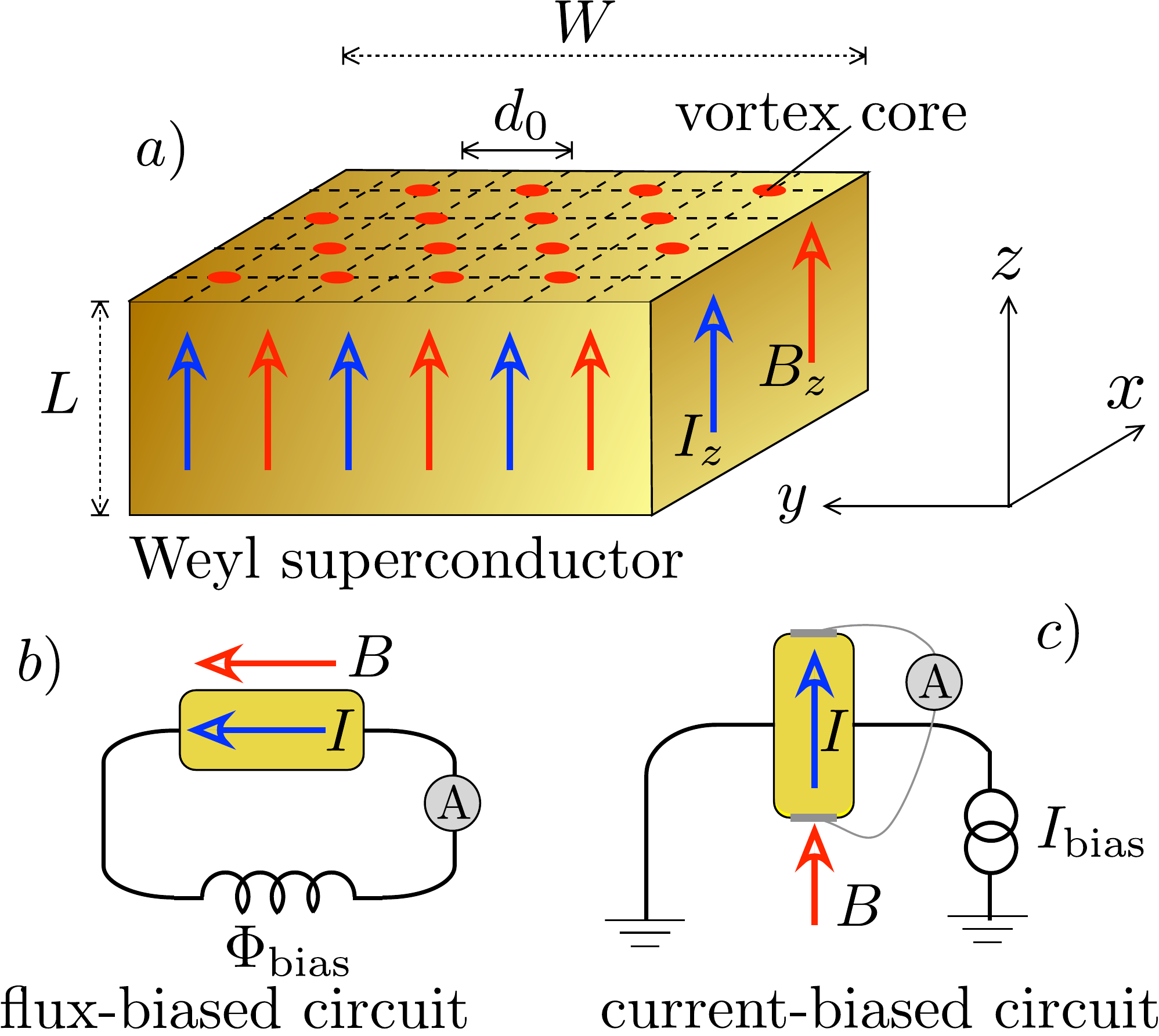}}
\caption{Panel a) shows a square vortex lattice in a Weyl superconductor, panels b) and c) show a circuit to measure the chiral magnetic effect (current $I$ parallel to an external magnetic field $B$). The current exists in equilibrium because Weyl fermions having one of the two chiralities are confined to vortex cores by a flux bias (panel b) or a current bias (panel c).
}
\label{fig_layout}
\end{figure}

As shown in Fig.\ \ref{fig_layout}, the heterostructure can be placed in either a flux-biased or a current-biased circuit. We seek the current $I_z$ in equilibrium, parallel to the external magnetic field $\bm{B}=\nabla\times\bm{A}$ in the $z$-direction.

The superconductor has length $L$ parallel to $B$, while the dimensions in the perpendicular direction are $W\times W$, large compared to the London penetration length $\lambda$. This is the key difference with Ref.\ \onlinecite{OBr17}, where $W<\lambda$ was assumed in order to prevent the formation of Abrikosov vortices. For $W\gg\lambda\gg l_m\gg \xi_0$ (with $l_m=\sqrt{\hbar/eB}$ the magnetic length and $\xi_0=\hbar v_{\rm F}/\Delta_0$ the superconducting coherence length) we are in the vortex phase of a strong-type-II superconductor, where the magnetic field penetrates in the form of vortices of magnetic flux $\Phi_0=h/2e$. The vortex lattice has two vortices per unit cell, we take the square array (lattice constant $d_0$) indicated in Fig.\ \ref{fig_layout}.

In the gauge with $\nabla\cdot\bm{A}=0$ the superconducting phase is determined by
\begin{equation}
\nabla\times\nabla\phi(\bm{r})=2\pi\hat{z}\sum_n\delta(\bm{r}-\bm{R}_n),\;\;\nabla\cdot\nabla\phi=0.\label{phidef}
\end{equation}
The first equation specifies a $2\pi$ winding of the phase around each vortex core at $\bm{R}_n$, and the second equation ensures that the superconducting velocity
\begin{equation}
m\bm{v}_s=\tfrac{1}{2}\nabla\phi-e\bm{A}\label{vsdef}
\end{equation}
has vanishing divergence. Since the vortex cores occupy only a small fraction $(\xi_0/l_m)^2$ of the volume, we may take a uniform pair potential amplitude $|\Delta|=\Delta_0$ and a uniform magnetic field strength $|\bm{B}|=B_0$. The dominant effect of the vortex lattice is the purely quantum mechanical scattering of quasiparticles by the superconducting phase \cite{Fra00}.

The vector potential contains a constant contribution $A_z=\Lambda/e$ in the $z$-direction controlled by either the flux bias or the current bias \cite{Tinkham}:
\begin{equation}
\Lambda=\begin{cases}
(e/L)\Phi_{\rm bias}&\text{(flux bias)},\\
e\mu_0(\lambda/W)^2 I_{\rm bias}&\text{(current bias)}.
\end{cases}\label{fluxcurrentbias}
\end{equation}

\section{Chirality confinement in a vortex lattice}
\label{sec_chiralitylocalization}

In the absence of a vortex lattice, for $W<\lambda$, it was shown in Ref.\ \onlinecite{OBr17} that a flux bias or current bias confines Weyl fermions of one definite chirality to the surfaces parallel to the magnetic field, gapping them out in the bulk. Here we consider the opposite regime $W\gg\lambda$ in which a vortex lattice forms in the Weyl superconductor. We will show that effect of the $\Lambda$ bias is qualitatively different: both chiralities remain gapless in the bulk, but one of the two chiralities is confined to the vortex cores.

The analytics is greatly simplified if the magnetic field is along the same $z$-axis as the separation of the Weyl cones. The corresponding vector potential is
\begin{equation}
\bm{A}(\bm{r})=(B_0y,0,\Lambda/e),\;\;\Lambda=(e/L)\Phi_{\rm bias},\label{Afluxbias}
\end{equation}
where for definiteness we take $\Lambda\geq 0$. This is the flux-biased geometry of Fig.\ \ref{fig_layout}b. Numerical simulations indicate that the current-biased geometry of Fig.\ \ref{fig_layout}c, with $\bm{B}$ along the $y$-axis, is qualitatively similar --- but we have not succeeded in obtaining a complete analytical treatment in that geometry.

\subsection{Landau bands}
\label{sec_numerics}

We have calculated the eigenvalues and eigenfunctions of the tight-binding Hamiltonian \eqref{Hamiltonian_def} using the \textit{Kwant} code \cite{kwant} as described in Ref.\ \onlinecite{Pac18}. We take parameters $\beta = t_0$, $\Delta_0=0.5\,t_0$, $\mu=0$. We arrange $h/2e$ vortices on the square lattice shown in Fig.\ \ref{fig_layout}a. The lattice constant $d_0=Na_0$ of the vortex lattice determines the magnetic field $B_0=(h/e)d_0^{-2}$. In the numerics the full nonlinear $\bm{k}$-dependence of ${\cal H}(\bm{k})$ is used, while for the analytical expressions we expand near $\bm{k}=0$.

\begin{figure*}[ht]
	\includegraphics[width=1\linewidth]{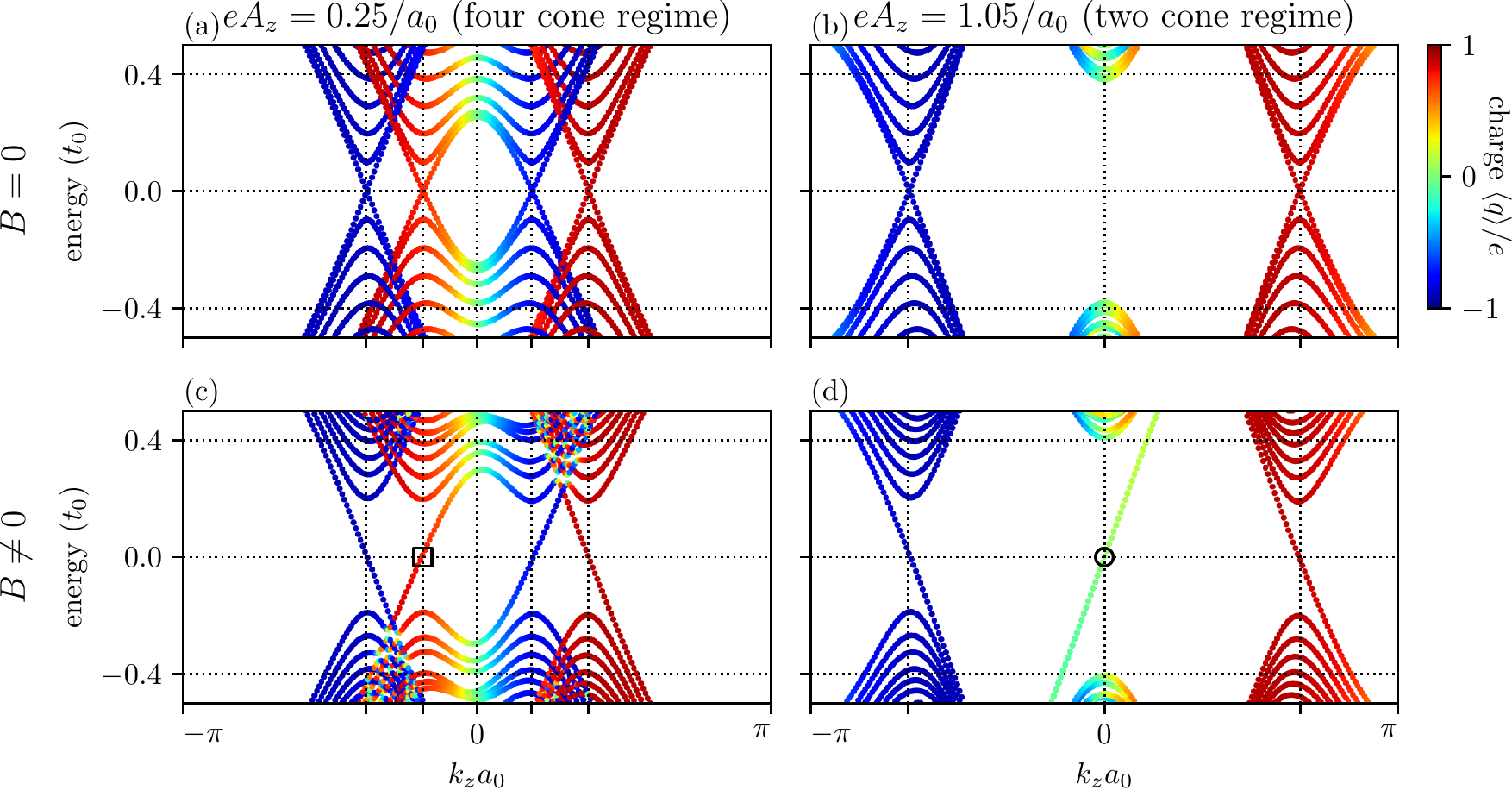}
	\caption{Dispersion relations of a Weyl superconductor at two values of the flux bias $A_z$ (left and right column), without and with a magnetic field $B$ (top and bottom row). In zero field and at a small value of the flux bias (panel \textit{a}), there are four cones in the spectrum. As the flux bias increases the electron-like cones (red) are shifted to positive $k_z$, whereas the hole-like cones (blue) are shifted to negative $k_z$. At the critical value $eA_z=\beta-\Delta_0=0.5/a_0$ two cones of opposite chirality meet at $k_z=0$, a gap opens and the system transitions into the two-cone regime (panel \textit{b}). When a magnetic field is applied, each Weyl cone gives rise to a chiral zeroth Landau level (panel \textit{c}). In the two-cone regime (panel \textit{d}) a pair of chiral Landau levels forms charge-neutral Majorana modes (green). The spectra were calculated for the tight-binding Hamiltonian \eqref{Hamiltonian_def}, with $\beta=t_0$, $\Delta=0.5t_0$, and $\mu=0$. The $B\neq 0$ data is for a square vortex lattice with lattice constant $d_0=18a_0$. For an electron-like Landau level marked with a square and for a Majorana mode marked with a circle we show the spatial probability density in Fig.\ \ref{fig:density}.
} \label{fig:spectra}
\end{figure*}

The zero-field spectra in Figs.\ \ref{fig:spectra}a and \ref{fig:spectra}b reproduce the findings of Ref.\ \onlinecite{OBr17}: For small $\Lambda$ and provided that $\Delta_0<\beta$ one sees two pairs of oppositely charged gapless Weyl cones, symmetrically arranged around $k_z=0$ at momenta $K_\pm$ and $-K_\pm$ given by
\begin{equation}
K_\pm=\sqrt{(\beta\pm\Lambda)^2-\Delta_0^2}.\label{kzdef}
\end{equation}
The pair at $|k_z|=K_-$ is displaced relative to the other pair at $|k_z|=K_+$ by the flux bias $\Lambda$, becoming gapped when $\Lambda$ is in the critical range
\begin{equation}
\Lambda\in (\beta-\Delta_0,\beta+\Delta_0)\equiv(\Lambda_{c1},\Lambda_{c2}).\label{criticalrange}
\end{equation}

Application of a magnetic field in Figs.\ \ref{fig:spectra}c and \ref{fig:spectra}d shows the formation of chiral zeroth-order Landau  bands: a pair of electron-like Landau levels of opposite chirality and a similar pair of hole-like Landau levels. The Landau bands have a linear dispersion in the $z$-direction, along the magnetic field, while they are dispersionless flat bands in the $x$--$y$ plane.

For $k_z$ near $K_\pm$ the electron-like and hole-like dispersions are given by \cite{Pac18}
\begin{subequations}
\label{LLdispersion}
\begin{equation}
\begin{split}
&E_{\rm electron}(\bm{k})=(-\mu-k_z+K_+)\cos\theta,\\
&E_{\rm hole}(\bm{k})=(\mu+k_z-K_-)\cos\theta,
\end{split}
\label{LLdispersiona}
\end{equation}
and similarly near $-K_\pm$ the dispersions are
\begin{equation}
\begin{split}
&E_{\rm electron}(\bm{k})=(-\mu+k_z+K_-)\cos\theta,\\
&E_{\rm hole}(\bm{k})=(\mu-k_z-K_+)\cos\theta.
\end{split}
\label{LLdispersionb}
\end{equation}
\end{subequations}
The $k_z$-dependent factor $\cos\theta$ renormalizes the charge and velocity of the quasiparticles, according to \cite{OBr17,Lem19}
\begin{equation}
\begin{split}
&\cos\theta(\bm{k})=\frac{|k_z|}{\sqrt{\Delta_0^2+k_z^2}}\\
&\quad\rightarrow \sqrt{1-\frac{\Delta_0^2}{(\beta\pm\Lambda)^2}}\equiv \kappa_\pm\;\;\text{when}\;\;|k_z|\rightarrow K_\pm.
\end{split}
\label{kappapmdef}
\end{equation}

The degeneracy of a Landau band is not affected by charge renormalization \cite{Pac18}, each electron-like or hole-like Landau band contains
\begin{equation}
{\cal N}_0=\tfrac{1}{2}\Phi/\Phi_0=(e/h)\Phi\label{calNdef}
\end{equation}
chiral modes, determined by the ratio of the enclosed flux $\Phi=B_0 W^2$ and the \textit{bare} single-electron flux quantum $h/e$.

\begin{figure}
	\includegraphics[width=1\linewidth]{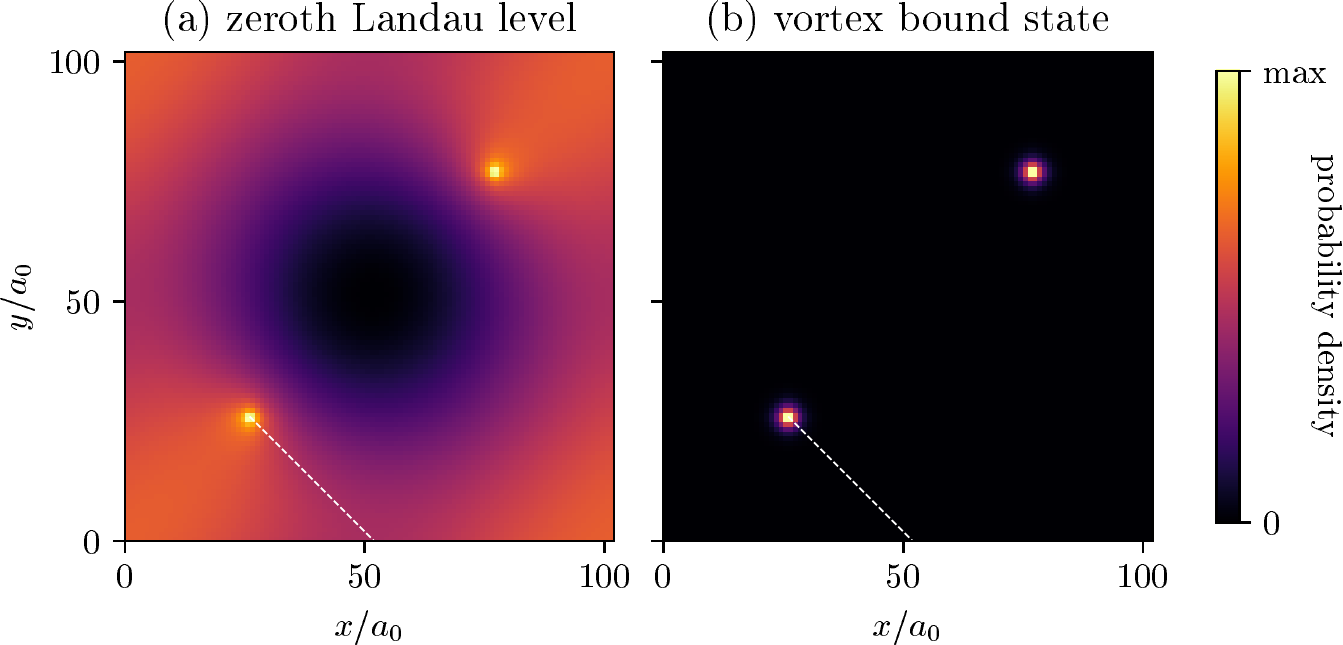}

	\vspace{0.2cm}
	\includegraphics[width=1\linewidth]{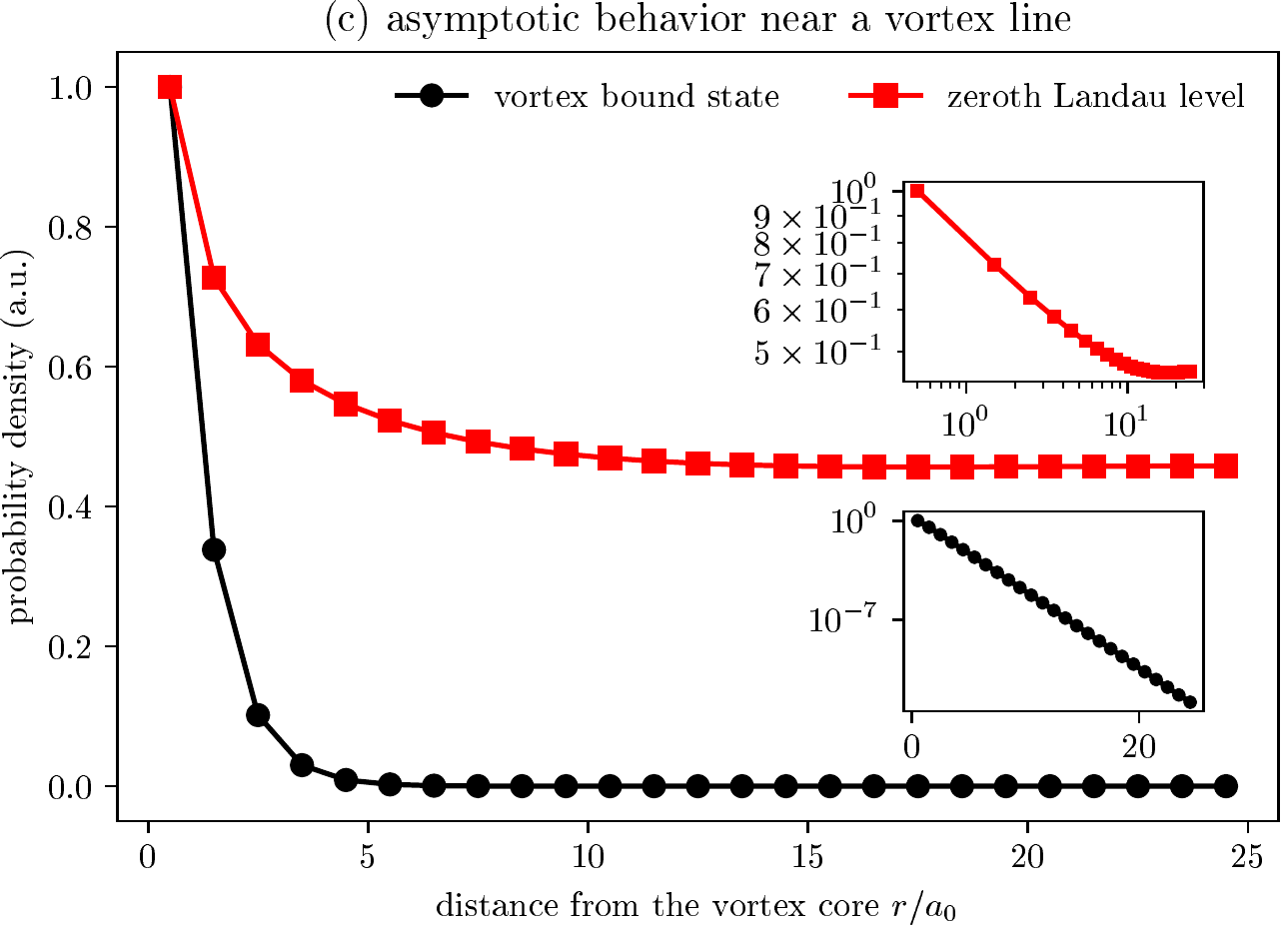}
	\caption{\label{fig:density}Spatial distribution of the probability density for an electron-like Landau band (panel \textit{a}) and for a Majorana vortex-core band (panel \textit{b}). Panel \textit{c} shows both probability distributions as a function of the distance $r$ from the vortex core, measured along the dashed white line in panels \textit{a,b}. In the insets in panel \textit{c} the same data is presented using a log-log scale (for the zeroth Landau level) and log-linear scale (for the vortex-core band). The Landau band is spread over the magnetic unit cell, with an \textit{algebraic} divergence at the vortex cores, whereas the vortex-core band is \textit{exponentially} localized at the vortices. The profiles were calculated for the same set of parameters as the spectra in Fig.\ \ref{fig:spectra}, with the Landau band corresponding to the state marked with a square, and the vortex-core band corresponding to the state marked with a circle. To improve the spatial resolution, we used a larger ratio $d_0/a_0=102$.}
\end{figure}

While the dispersion of a Landau band in the Brillouin zone changes only quantitatively with the flux bias, it does have a pronounced qualitative effect on the \textit{spatial} extension in the $x$--$y$ plane. As shown in Fig.\ \ref{fig:density}, the intensity profile $|\psi_{\pm}(x,y)|^2$ of a zeroth-order Landau level at $|k_z|=K_\pm$ peaks when $\bm{r}=(x,y)$ approaches a vortex core at $\bm{R}_n$. The dependence on the separation $\delta r=|\bm{r}-\bm{R}_n|$ is a power law \cite{Pac18},
\begin{equation}
|\psi_{\pm}|^2\propto \delta r^{-1+\kappa_\pm}.\label{kappadef}
\end{equation}

When $\Lambda$ enters the critical range \eqref{criticalrange} this power law decay applies only to one of the two chiralities: the two Landau bands at $k_z=K_+$ and $k_z=-K_+$ with $dE/dk_z<0$ still have the power law decay \eqref{kappadef}, but the other two bands with $dE/dk_z>0$ merge at $k_z=0$ and become \textit{exponentially confined} to a vortex core. As we shall derive in the next subsection,
\begin{equation}
\begin{split}
&|\psi_{\rm vortex}|^2\propto \exp(-\delta r/l_{\rm conf}),\\
&l_{\rm conf}=\tfrac{1}{2}\max\left(\frac{1}{\Lambda-\beta+\Delta_0},\frac{1}{\beta-\Lambda+\Delta_0}\right).
\end{split}
\label{expdecay}
\end{equation}
These two vortex-core bands are separated spatially, one in each of the two vortices in the unit cell. They form unpaired Majorana fermions, in contrast to the two Landau bands that overlap spatially and as a pair constitute a Dirac fermion.

All of this applies to magnetic fields in the regime $W\gg\lambda\gg l_m\gg\xi_0$ of a vortex lattice. At weaker fields, when $l_m\gtrsim\min(W,\lambda)$, no vortices can form and the analysis of Ref.\ \onlinecite{Pac18} applies: The bands with chirality $dE/dk_z>0$ are pushed out of the bulk and confined to the surfaces along the $z$-direction.

\subsection{Vortex core bands}
\label{sec_analytics}

To demonstrate the exponential confinement in a vortex core of the $\tau_z=+1$ chirality we expand the Hamiltonian \eqref{Hamiltonian_def} to first order in $k_x,k_y$ at $k_z=0$, $\mu=0$,
\begin{align}
{\cal H} ={}&
\begin{pmatrix}
k_x\sigma_x+k_y\sigma_y &0\\
0&-k_x\sigma_x-k_y\sigma_y
\end{pmatrix}\nonumber\\
&+\begin{pmatrix}
(\beta-\Lambda)\sigma_z& \Delta_0 {e}^{{i}\varphi} \\ \Delta_0 {e}^{-{i}\varphi} &
(\beta-\Lambda)\sigma_z
\end{pmatrix}.\label{Hnearvortex}
\end{align}
The applied magnetic field does not contribute on length scales below $l_m$, so we only need to include the constant $eA_z=\Lambda$ term in the vector potential. The winding of the superconducting phase is accounted for by the factor $e^{i\varphi}$, in polar coordinates $(x,y,z)=(r\cos\varphi,r\sin\varphi,z)$ centered on the vortex core.

In view of the identity
\begin{equation}
\partial_x+i\partial_y=e^{i\varphi}\left(\partial_r+ir^{-1}\partial_\varphi\right),\label{polarcoorda}
\end{equation}
with $\partial_q\equiv\partial/\partial q$, the Hamiltonian \eqref{Hnearvortex} reads
 \begin{subequations}
 \label{Hpolardef}
\begin{align}
&{\cal H} = \begin{pmatrix}
(\beta-\Lambda)\sigma_z-D& \Delta_0 {e}^{{i}\varphi} \\ \Delta_0 {e}^{-{i}\varphi} &
(\beta-\Lambda)\sigma_z+D
\end{pmatrix},\label{Hpolar}\\
&D =  \begin{pmatrix}
0 & {e}^{-{i}\varphi}(i\partial_r +r^{-1}\partial_\varphi)\\
{e}^{{i}\varphi} (i\partial_r -r^{-1}\partial_\varphi)& 0
\end{pmatrix}.\label{sigmarphidef}
\end{align}
\end{subequations}

We seek a solution ${\cal H}\Psi=0$ of the form
\begin{equation}
\Psi=\bigl(\phi_1(r),
e^{i\varphi}\phi_2(r),
e^{-i\varphi}\phi_3(r),
\phi_4(r)\bigr),\label{psiansatz}
\end{equation}
and denote $\Phi=(\phi_1,\phi_2,\phi_3,\phi_4)$. This produces the ordinary differential equation
\begin{align}
-\frac{d\Phi}{dr}&= \begin{pmatrix}
0 & -{i}(\beta-\Lambda) & 0 & {i}\Delta_0 \\
{i}(\beta-\Lambda) & r^{-1} & {i}\Delta_0 & 0\\
0 & -{i}\Delta_0 & r^{-1} & {i}(\beta-\Lambda) \\
-{i}\Delta_0 & 0 & -{i}(\beta-\Lambda) & 0
\end{pmatrix}\Phi \nonumber\\
&\equiv \left(M_1 + r^{-1}{M_2}\right)\Phi.\label{psimatrixeq}
\end{align}

In the critical regime $\Lambda_{c1}<\Lambda<\Lambda_{c2}$ the two positive eigenvalues of the matrix $M_1$ are $\Lambda-\Lambda_{c1}$ and $\Lambda_{c2}-\Lambda$. At large $r$, the normalizable solution of Eq.\ \eqref{psimatrixeq} decays $\propto{e}^{-\alpha r}$, with $\alpha$ the smallest positive eigenvalue of $M_1$:
\begin{equation}
\alpha= \min(\Lambda-\Lambda_{c1},\Lambda_{c2}-\Lambda).\label{lconfresult}
\end{equation}
The confinement length $l_{\rm conf}=1/2\alpha$ is thus given by Eq.\ \eqref{expdecay}.

\section{Chiral magnetic effect}
\label{sec_CME}

\subsection{Charge renormalization}
\label{sec_eqcurrentdispersion}

We summarize the formulas from Ref.\ \onlinecite{OBr17} that show how charge renormalization by the superconductor affects the CME.

The equilibrium expectation value $I_{z}$ of the electrical current in the $z$-direction is given by
\begin{equation}
I_{z}=\tfrac{1}{2}\sum_{n}\int\frac{dk_z}{2\pi}\,f(E)\langle j_z\rangle_E.\label{Izdefapp2}
\end{equation}
The sum over $n$ is over transverse modes with energy $E_n(k_z)\equiv E$ at longitudinal momentum $k_z$, weighted by the Fermi function $f(E)=(1+e^{E/k_{\rm B}T})^{-1}$ at temperature $T$. The factor $1/2$ corrects for a double-counting of states in the Bogoliubov-De Gennes formalism. The expectation value of the current operator $j_z= -\partial H/\partial A_z$ in the state with energy $E$ equals
\begin{equation}
\langle j_z\rangle_E=-\langle\partial H/\partial A_z\rangle_E=-\partial E/\partial A_z,\label{jzaverage}
\end{equation}
according to the Hellmann-Feynman theorem. Two other expectation values that we need are those of the velocity operator $v_z=\partial H/\partial k_z$ and the charge operator $Q=-e\partial H/\partial \mu$, given by
\begin{equation}
\langle v_z\rangle_E=\partial E/\partial k_z,\;\;\langle Q\rangle_E=-e\partial E/\partial \mu.\label{vzQaverage}
\end{equation}
Following Ref. \onlinecite{OBr17} we also define the ``vector charge''
\begin{equation}
\bm{Q}=(Q_x,Q_y,Q_z),\;\;\text{with}\;\;Q_\alpha(E) \equiv \frac{\langle j_\alpha\rangle_E}{\langle v_\alpha\rangle_E},
\end{equation}
which may be different from the average (scalar) charge $Q_0\equiv\langle Q\rangle_E$ because the average of the current as the product of charge and velocity may differ from the product of the averages.

The CME is a contribution to $I_z$ that is linear in the equilibrium chemical potential $\mu$, measured relative to the Weyl points. We extract this contribution by taking the derivative $\partial_\mu I_z$ in the limit $\mu\rightarrow 0$. Two terms appear, an \textit{on-shell} term from the Fermi level and an \textit{off-shell} term from energies below the Fermi level,
\begin{subequations}
\label{Joffshellonshell}
\begin{align}
\partial_\mu I_z={}&{\cal J}_{\text{on-shell}}+{\cal J}_{\text{off-shell}} \equiv \mathcal{J}_\text{total},\label{dIalphadmu}\\
{\cal J}_{\text{on-shell}}={}&-\frac{1}{2e}\sum_{n}\int\frac{dk_z}{2\pi}f'(E)\langle Q\rangle_E\langle j_z\rangle_E,\label{Jonshell}\\
{\cal J}_{\text{off-shell}}={}&-\frac{1}{2}\sum_{n}\int\frac{dk_z}{2\pi}f(E)\frac{\partial^2}{\partial
A_z\partial\mu}E_n(k_z).\label{Joffshell}
\end{align}
\end{subequations}

At low temperatures, when $-f'(E)\rightarrow \delta(E)$ becomes a delta function, the on-shell contribution ${\cal J}_{\text{on-shell}}$ involves only Fermi surface properties. It is helpful to rewrite it as a sum over modes at $E=0$. For that purpose we replace the integration over $k_z$ by an energy integration weighted with the density of states:
\begin{equation}
{\cal J}_{\text{on-shell}}=-\frac{1}{4\pi e}\sum_{n}\int_{-\infty}^\infty dE\, f'(E)\left|\frac{\partial E}{\partial k_z}\right|^{-1}
\langle Q\rangle_E\langle j_z\rangle_E.\label{IalphaFermileveldE}
\end{equation}
In the $T\rightarrow 0$ limit a sum over modes remains,
\begin{equation}
{\cal J}_{\text{on-shell}}=\frac{1}{2}\frac{e}{h}\sum_{n}
\frac{Q_0 Q_z}{e^2}\,\bigl({\rm sign}\,\langle v_z\rangle\bigr)\biggr|_{E_{n}=0},\label{IalphaFermilevel}
\end{equation}
where we have restored the units of $\hbar=h/2\pi$.

\subsection{On-shell contributions}
\label{sec_onshell}

We apply Eq.\ \eqref{IalphaFermilevel} to the vortex lattice of the flux-biased Weyl superconductor. Derivatives with respect to $A_z$ are then derivatives with respect to the flux bias $\Lambda$. According to the dispersion relation \eqref{LLdispersiona}, the electron-like Landau band near $K_+$ has renormalized charges
\begin{equation}
Q_0=e\kappa_+,\;\;Q_z=e\frac{\partial K_+}{\partial \Lambda}=\frac{e}{\kappa_+},\label{chargeelectronplus}
\end{equation}
in the limit $k_z\rightarrow K_+$, $\mu\rightarrow 0$. The charge renormalization factors cancel, so this Landau band with ${\rm sign}\,\langle v_z\rangle <0$ contributes to ${\cal J}_{\text{on-shell}}$ an amount $-\frac{1}{2}e/h$ times the degeneracy ${\cal N}_0=(e/h)\Phi$, totalling $-\frac{1}{2}(e/h)^2\Phi$.

Similarly, for the hole-like Landau band near $-K_-$ Eq.\ \eqref{LLdispersiona} gives
\begin{equation}
Q_0=-e\kappa_+,\;\;Q_z=-e\frac{\partial K_+}{\partial \Lambda}=-\frac{e}{\kappa_+},\label{chargeforplus}
\end{equation}
for the same contribution of $-\frac{1}{2}(e/h)^2\Phi$. The total on-shell contribution for this chirality is
\begin{equation}
{\cal J}_{\text{on-shell}}(|k_z|=K_+)=-(e/h)^2\Phi.\label{JonshellKplus}
\end{equation}

We can repeat the calculation for the electron-like band near $K_-$ and the hole-like band near $-K_-$, the only change is the ${\rm sign}\,\langle v_z\rangle >0$, resulting in
\begin{equation}
{\cal J}_{\text{on-shell}}(|k_z|=K_-)=(e/h)^2\Phi.\label{JonshellKplus}
\end{equation}

We conclude that the Dirac fermions in the Landau bands of opposite chirality give identical opposite on-shell contributions $\pm(e/h)^2\Phi$ to $\partial_\mu I_z$. The net result vanishes when $\Lambda$ is outside of the critical region $(\Lambda_{c1},\Lambda_{c2})$. When $\Lambda_{c1}<\Lambda<\Lambda_{c2}$ one of the two chiralities is transformed into unpaired Majorana fermions confined to the vortex cores. The vortex-core bands have $Q_0=0$ at $E=0$, so they have no on-shell contribution, resulting in
\begin{equation}
{\cal J}_{\text{on-shell}}=\begin{cases}
0&\text{if}\;\;\Lambda\notin(\Lambda_{c1},\Lambda_{c2}),\\
(e/h)^2\Phi&\text{if}\;\;\Lambda\in(\Lambda_{c1},\Lambda_{c2}).
\end{cases}\label{totalonshell}
\end{equation}
The coefficient $(e/h)^2$ contains the bare charge, unaffected by the charge renormalization.

\begin{figure}[htb!]
	\includegraphics[width=1\linewidth]{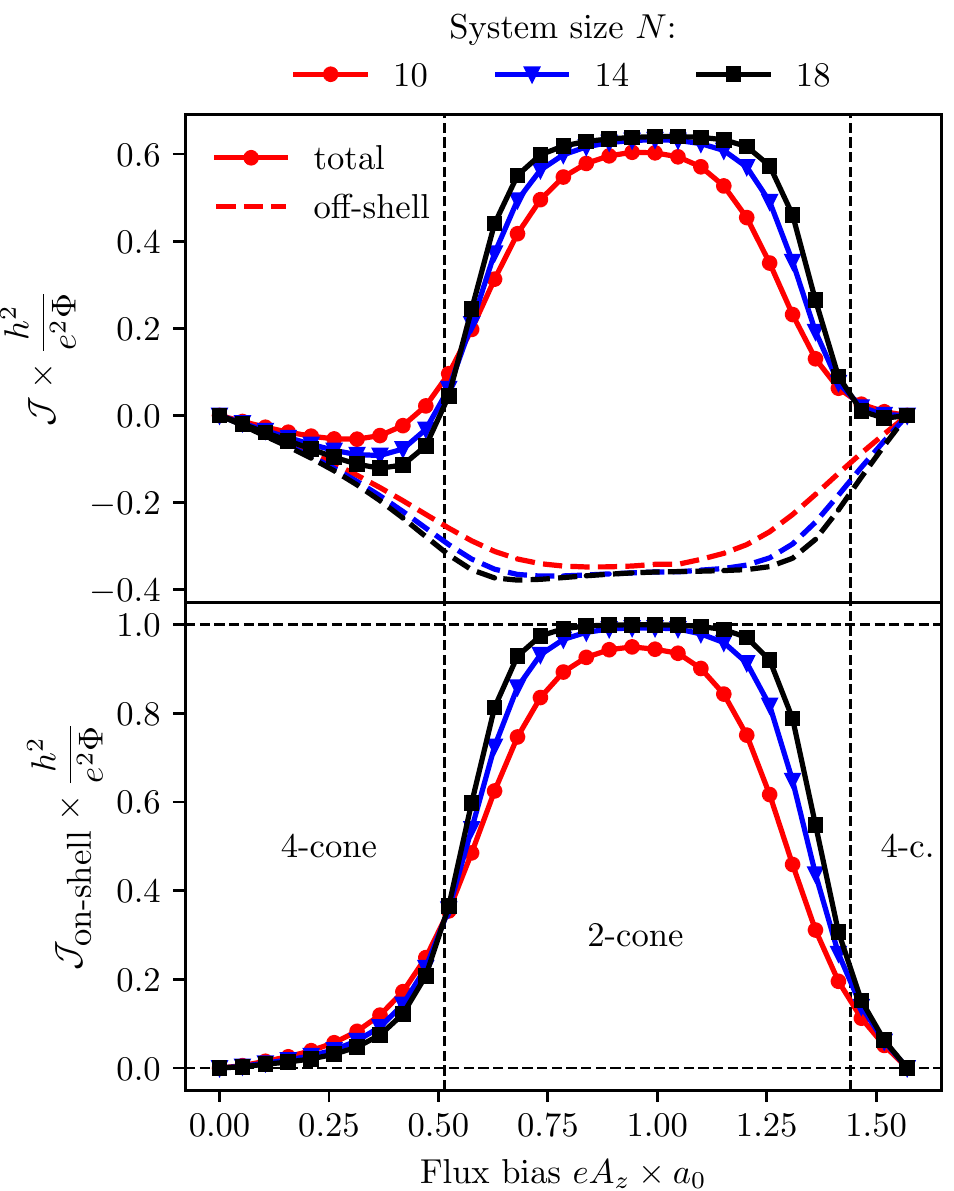}
	\caption{Numerical calculation of $\partial_\mu I_z$ at $\mu=0.05\,t$ in the tight-binding Hamiltonian \eqref{Hamiltonian_def}. The solid curves are the total current, while the dashed curves show only the off-shell contribution \eqref{Joffshell}.  The vertical dashed lines mark $eA_{z}=\Lambda_{c1},\Lambda_{c2}$ -- the values of the flux bias which correspond to a topological phase transition into and out of the two-cone regime. The horizontal dashed lines mark the universal CME value of $(h/e)^2\Phi$. As the size $N=d/a_0$ of the magnetic unit cell increases, the numerically calculated value of the on-shell contribution approaches the universal value, which jumps at the topological phase transition. \label{fig:phase_transition}}
\end{figure}

\begin{figure}[htb!]
	\includegraphics[width=0.9\linewidth]{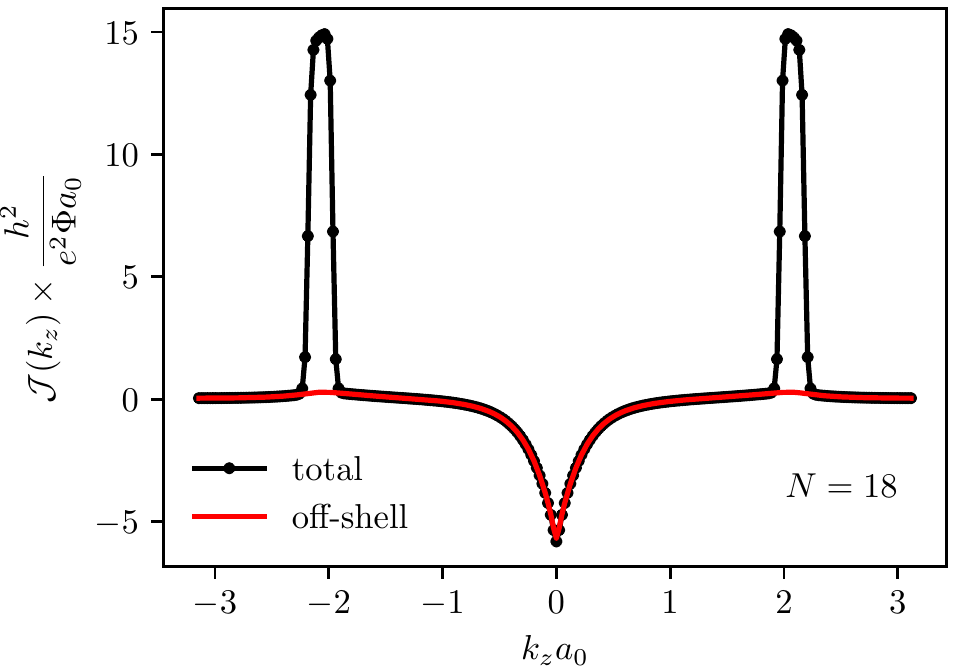}
	\caption{Same numerical calculation as in Fig.\ \ref{fig:phase_transition}, but now for a fixed flux bias $eA_z = 1.05/a_0$ in the two-cone regime, showing the contributions to $\partial_\mu I_z$ from different momenta $k_z$ along the magnetic field. We distinguish between the total current and the off-shell contribution. The difference between the two is the on-shell contribution, which peaks at the momenta where the Fermi level crosses the chiral Landau bands. The vortex-core bands at $k_z=0$ have vanishing on-shell contribution. \label{fig:function_of_kz}}
\end{figure}

\subsection{Off-shell contributions}
\label{sec_offshell}

Turning now to the off-shell contributions \eqref{Joffshell}, we note that the Landau bands do not contribute in view of Eq.\ \eqref{LLdispersion}:
\begin{equation}
\frac{\partial^2}{\partial\Lambda\partial\mu}E(\bm{k})=\pm \frac{\partial}{\partial\Lambda}\cos\theta(\bm{k})=0.
\end{equation}
For the vortex-core bands, off-shell contributions cancel because of particle-hole symmetry.

This does not exclude off-shell contributions from states far below the Fermi level, where our entire low-energy analysis no longer applies. In fact, as we show in Figs.\ \ref{fig:phase_transition} and \ref{fig:function_of_kz}, we do find a substantial off-shell contribution to $\partial_\mu I_z$ in our numerical calculations (see App.\ \ref{sec_numerics} for details). Unlike the on-shell contribution \eqref{totalonshell}, which has a discontinuity at $\Lambda=\Lambda_{c1},\Lambda_{c2}$, the off-shell contribution depends smoothly on the flux bias and can therefore be extracted from the data.

\section{Conclusion}
\label{sec:conclude}

In summary, we have demonstrated that a flux bias in a Weyl superconductor drives a confinement/deconfinement transition in the vortex phase: For weak flux bias the subgap excitations are all delocalized in the plane perpendicular to the vortices. With increasing flux bias a transition occurs at which half of the states become exponentially localized inside the vortex cores. The localized states have a definite chirality, meaning that they all propagate in the same direction along the vortices. (The sign of the velocity is set by the sign of the external magnetic field $B_0$.)

As a physical consequence of this topological phase transition we have studied the chiral magnetic effect. The states confined to the vortex cores are charge-neutral Majorana fermions, so they carry no electrical current. The states of opposite chirality, which remain delocalized, are charged, and because they all move in the same direction they can carry a nonzero current density $j$ parallel to the vortices. This is an equilibrium supercurrent, proportional to the magnetic field $B_0$ and to the chemical potential $\mu$ (measured relative to the Weyl point). 

We have calculated that the supercurrent along the vortices jumps at the topological phase transition by an amount which for a large system size tends to the universal limit
\begin{equation}
j=\frac{e^2}{h^2}\,B_0\mu.
\end{equation}
Remarkably enough, the proportionality constant contains the bare electron charge $e$, even though the quasiparticles have a renormalized charge $e^\ast<e$. This electromagnetic response is generated by the axion term $(e/h)^2\int dt \int d\bm{r} \,\theta(t)\, E_z B_z $ in the Lagrangian, where $\theta(t)=\mu t$ is the axion angle.

The chiral fermions confined in the vortex cores are a superconducting realization of the ``topological coaxial cable'' of Schuster \textit{et al.} \cite{Sch16}, where the fermions are confined to vortex lines in a Higgs field. There is one difference: the chiral fermions in the Higgs field are charge-$e$ Dirac fermions, while in our case they are charge-neutral Majorana fermions. The difference manifests itself in the physical observable that serves as a signature of the confinement: for Schuster \textit{et al.} this is a quantized current $dI/dV=e^2/h$ per vortex out of equilibrium, in our case it is a quantized current $dI/d\mu =\tfrac{1}{2}e/h$ per vortex in equilibrium.

\acknowledgments

This project has received funding from the Netherlands Organization for Scientific Research (NWO/OCW) and from the European Research Council (ERC) under the European Union's Horizon 2020 research and innovation programme.

\appendix

\section{Details of the numerical calculation}
\label{sec_numerics}

The numerical calculation was performed on a square lattice with two $h/2e$ vortices in a magnetic unit cell, using the discretization described in Ref. \onlinecite{Pac18}. We calculate separately the total induced current response
\begin{equation}
\partial_\mu I_z={\cal J}_{\text{on-shell}}+{\cal J}_{\text{off-shell}}\equiv {\cal J}_{\text{total}},
\end{equation}
and the off-shell contribution ${\cal J}_{\text{off-shell}}$.  The defining equations \eqref{Izdefapp2} and \eqref{Joffshell} are rewritten in terms of finite differences,
\begin{widetext}
	\begin{align}
	{\cal J}_{\text{total}}&=\frac{1}{2}\lim_{\mu\rightarrow 0}\frac{1}{2\mu}  \sum_n \int\frac{dk_z}{2\pi}\, \biggl[f\bigl(E_n(k_z, \mu)\bigr) \langle j_z\rangle_{E_n(k_z,\mu)}  - f\bigl(E_n(k_z, -\mu)\bigr) \langle j_z\rangle_{E_n(k_z,-\mu)}\biggr] \nonumber \\
	&\qquad = \frac{1}{2}\lim_{\mu\rightarrow 0}\frac{1}{2\mu} \int dk_z\,  \Delta I_z^{\text{total}}(\mu, k_z) = \lim_{\mu\rightarrow 0}\frac{1}{\mu} \Delta I_z^{\text{total}}(\mu), \label{dIz_tot}
	\\
	{\cal J}_{\text{off-shell}} &= \frac{1}{2}
	\lim_{\mu\rightarrow 0} \frac{1}{2\mu}\sum_n \int\frac{dk_z}{2\pi}\,f\bigl(E_n(k_z, \mu=0)\bigr)  \biggl[\langle j_z\rangle_{E_n(k_z,\mu)}-\langle j_z\rangle_{E_n(k_z,-\mu)}\biggr]
	\nonumber \\
	&\qquad = \frac{1}{2}\lim_{\mu\rightarrow 0}\frac{1}{2\mu} \int dk_z\,  \Delta I_z^{\text{off-shell}}(\mu, k_z) = \lim_{\mu\rightarrow 0}\frac{1}{\mu} \Delta I_z^{\text{off-shell}}(\mu). \label{dIz_os}
	\end{align}
\end{widetext}

\begin{figure*}[tb]
	\centerline{\includegraphics[width=0.9\linewidth]{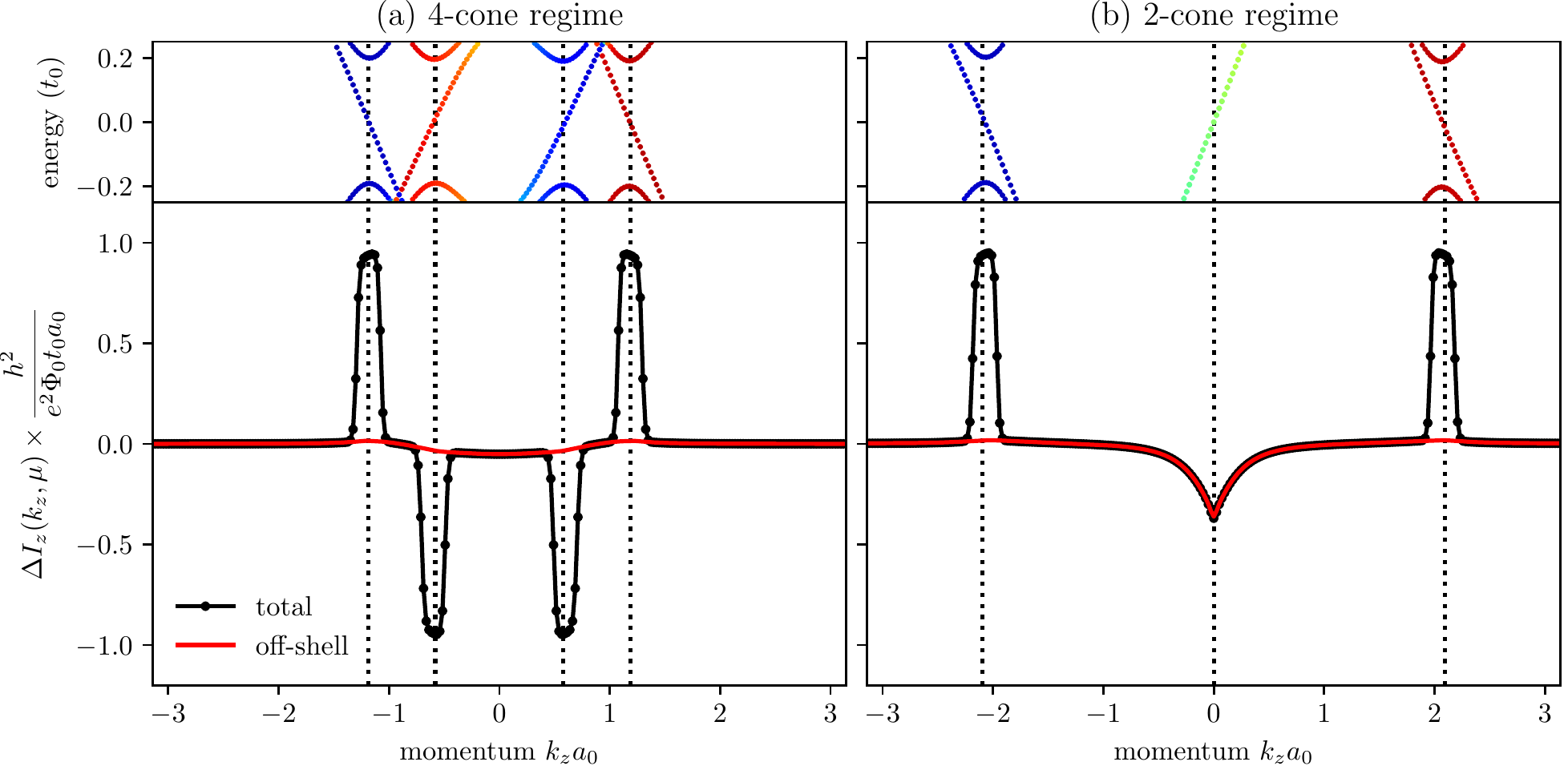}}
	\caption{Bottom: momentum-resolved current response $\Delta I_z(\mu, k_z)$, as defined in Eqs.\ \eqref{dIz_tot} and \eqref{dIz_os}, in the four-cone regime at $eA_z=0.25/a_0$ (panel \textit{a}) and in the two-cone regime at $eA_z = 1.05/a_0$ (panel \textit{b}). Top: low-energy dispersion relation for the corresponding system. The on-shell contribution to the current response, which is the difference between the total and off-shell contributions, only appears at momenta for which a band crosses the Fermi energy. In the four-cone regime four peaks are present, the contributions of which cancel out. In the two-cone regime the vortex-core band at $k_z=0$ has a vanishing on-shell contribution, whereas the contribution of the other two Landau levels remains unchanged. The plots were obtained for a system size $N=18$.}
	\label{function_of_kz_ab}
\end{figure*}

\begin{figure}[tb]
	\centerline{\includegraphics[width=0.9\linewidth]{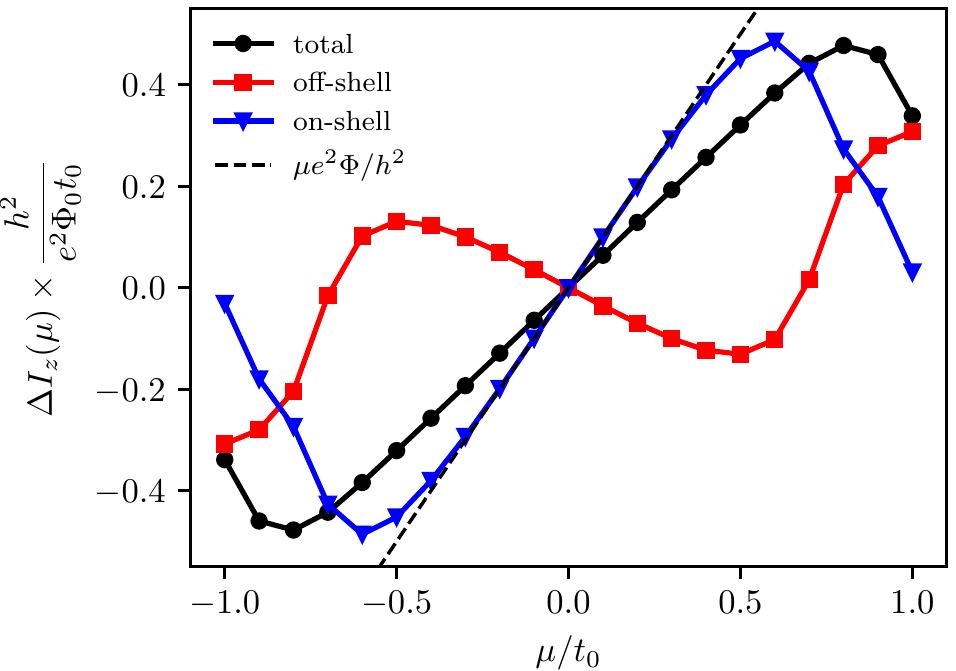}}
	\caption{
		The current response $\Delta I_z(\mu)$, as defined in Eqs.\ \eqref{dIz_tot} and \eqref{dIz_os}, in the two-cone regime at $eA_z=1.05/a_0$ for a finite chemical potential $\mu$. The colored data points give the total response, as well as the off-shell and on-shell contributions. The dotted line $\mu e^2 \Phi/h^2$ is the theoretical prediction \eqref{totalonshell} for the on-shell contribution to first order in $\mu$, which is a good approximation to the numerical result for small $\mu$. The plots were obtained for a system size $N=18$.}
	\label{function_of_mu}
\end{figure}

We computed the values of the expressions on the right-hand-side at finite $\mu$. The $k_z$-integral was estimated from 256 values of $k_z$, equally spaced in the $[-\pi, \pi]$ interval. For the sum over transverse modes $n$ we averaged over 4 values of both $k_x$ and $k_y$. To smoothen the integrand we took a small nonzero temperature $T=0.01$ in the Fermi function --- much smaller than the energy of the first Landau level (which was $\gtrsim 0.2$ for the parameters we considered). In Fig.\ \ref{function_of_kz_ab} we present the results prior to integration over $k_z$, for two different values of $A_z$. For $\mu=0.05$ the finite differences have converged to the derivative -- see Fig.\ \ref{function_of_mu}.


\begin{thebibliography}{99}
\bibitem{Vaz13} M. M. Vazifeh and M. Franz, \textit{Electromagnetic response of Weyl semimetals}, Phys. Rev. Lett. \textbf{111}, 027201 (2013).
\bibitem{Che13} Y. Chen, S. Wu, and A. A. Burkov, \textit{Axion response in Weyl semimetals}, Phys. Rev. B \textbf{88}, 125105 (2013).
\bibitem{Zho13} Jian-Hui Zhou, Hua Jiang, Qian Niu, and Jun-Ren Shi, \textit{Topological invariants of metals and the related physical effects}, Chin. Phys. Lett. \textbf{30}, 027101 (2013).
\bibitem{Ma15} J. Ma and D. A. Pesin, \textit{Chiral magnetic effect and natural optical activity in metals with or without Weyl points}, Phys. Rev. B \textbf{92}, 235205 (2015).
\bibitem{Ala16} Y. Alavirad and J. D. Sau, \textit{Role of boundary conditions, topology, and disorder in the chiral magnetic effect in Weyl semimetals}, Phys. Rev. B \textbf{94}, 115160 (2016).
\bibitem{Zho16a} S. Zhong, J. E. Moore, and I. Souza, \textit{Gyrotropic magnetic effect and the magnetic moment on the Fermi surface}, Phys. Rev. Lett. \textbf{116}, 077201 (2016).
\bibitem{Bai16a} P. Baireuther, J. A. Hutasoit, J Tworzyd{\l}o, and C. W. J. Beenakker, \textit{Scattering theory of the chiral magnetic effect in a Weyl semimetal: interplay of bulk Weyl cones and surface Fermi arcs}, New J. Phys. \textbf{18}, 045009 (2016).
\bibitem{Zub16} M. A. Zubkov, \textit{Absence of equilibrium chiral magnetic effect}, Phys. Rev. D \textbf{93}, 105036 (2016).
\bibitem{Vol17} G. E. Volovik, \textit{Chiral vortical effect generated by chiral anomaly in vortex-skyrmions}, JETP Lett. \textbf{105}, 303 (2017).
\bibitem{OBr17} T. E. O'Brien, C. W. J. Beenakker, and I. Adagideli, \textit{Superconductivity provides access to the chiral magnetic effect of an unpaired Weyl cone}, Phys. Rev. Lett. \textbf{118}, 207701 (2017).
\bibitem{Men19} T. Meng and J. C. Budich, \textit{Unpaired Weyl nodes from long-ranged interactions: Fate of quantum anomalies}, Phys. Rev. Lett. \textbf{122}, 046402 (2019).
\bibitem{Mas17} G. Massarelli, G. Wachtel, J. Y. T. Wei, and A. Paramekanti, \textit{Pseudo-Landau levels of Bogoliubov quasiparticles in nodal superconductors}, Phys. Rev. B \textbf{96}, 224516 (2017).
\bibitem{Liu17} T. Liu, M. Franz, and S. Fujimoto, \textit{Quantum oscillations and Dirac-Landau levels in Weyl superconductors}, Phys. Rev. B \textbf{96}, 224518 (2017).
\bibitem{Nic17} E. M. Nica and M. Franz, \textit{Landau levels from neutral Bogoliubov particles in two-dimensional nodal superconductors under strain and doping gradients}, Phys. Rev. B \textbf{97}, 024520 (2018).
\bibitem{Pac18} M. J. Pacholski, C. W. J. Beenakker, and I. Adagideli, \textit{Topologically protected Landau level in the vortex lattice of a Weyl superconductor}, Phys. Rev. Lett. \textbf{121}, 037701 (2018).
\bibitem{Nie83} H. B. Nielsen and M. Ninomiya, \textit{The Adler-Bell-Jackiw anomaly and Weyl fermions in a crystal}, Nucl. Phys. B \textbf{130}, 389 (1983).
\bibitem{Kha14} D. Kharzeev, \textit{The Chiral Magnetic Effect and anomaly-induced transport}, Progr. Part. Nucl. Phys. \textbf{75}, 133 (2014).
\bibitem{Bur15} A. A. Burkov, \textit{Chiral anomaly and transport in Weyl metals}, J. Phys. Condens. Matter \textbf{27}, 113201 (2015).
\bibitem{Lev85} L. S. Levitov, Yu. V. Nazarov, and G. M. Eliashberg, \textit{Magnetostatics of superconductors without an inversion center}, JETP Lett. \textbf{41}, 445 (1985).
\bibitem{Naz86} Yu. V. Nazarov, \textit{Instability due to magnetically induced currents}, Sov. Phys. JETP \textbf{64}, 193 (1986).
\bibitem{Gor98} L. P. Gor'kov and J. R. Schrieffer, \textit{de Haas--van Alphen effect in anisotropic superconductors in magnetic fields well below $H_{c2}$}, Phys. Rev. Lett. \textbf{80}, 3360 (1998).
\bibitem{And98} P. W. Anderson, \textit{Anomalous magnetothermal resistance of high-$T_c$ superconductors: Anomalous cyclotron orbits at a Dirac point}, arXiv:cond-mat/9812063.
\bibitem{Mel99} A. S. Mel'nikov, \textit{Quantization of the quasiparticle spectrum in the mixed state of d-wave superconductors}, J. Phys. Condens. Matter \textbf{11}, 4219 (1999).
\bibitem{Fra00} M. Franz and Z. Te\v{s}anovi\'{c}, \textit{Quasiparticles in the vortex lattice of unconventional superconductors: Bloch waves or Landau levels?}, Phys. Rev. Lett. \textbf{84}, 554 (2000).
\bibitem{Mar00} L. Marinelli, B. I. Halperin, and S. H. Simon, \textit{Quasiparticle spectrum of d-wave superconductors in the mixed state}, Phys. Rev. B \textbf{62}, 3488 (2000).
\bibitem{Bur11} A. A. Burkov and L. Balents, \textit{Weyl semimetal in a topological insulator multilayer}, Phys. Rev. Lett. \textbf{107}, 127205 (2011).
\bibitem{Men12} T. Meng and L. Balents, \textit{Weyl superconductors}, Phys. Rev. B \textbf{86}, 054504 (2012). \textit{Erratum:} Phys. Rev. B \textbf{96}, 019901 (2017).
\bibitem{Tinkham} M. Tinkham, \textit{Introduction to Superconductivity} (Dover Publications, 2004).
\bibitem{Lem19} G. Lemut, M. J. Pacholski, I. Adagideli, and C. W. J. Beenakker, \textit{Effect of charge renormalization on electric and thermo-electric transport along the vortex lattice of a Weyl superconductor}, Phys. Rev. B \textbf{100}, 035417 (2019).
\bibitem{kwant} C. W. Groth, M. Wimmer, A. R. Akhmerov, and X. Waintal, \textit{Kwant: A software package for quantum transport}, New J. Phys. \textbf{16}, 063065 (2014).
\bibitem{Sch16} T. Schuster, T. Iadecola, C. Chamon, R. Jackiw, S.-Y. Pi, \textit{Dissipationless conductance in a topological coaxial cable}, Phys. Rev. B \textbf{94}, 115110 (2016).
\end{thebibliography}
\end{document}